# Microscale Heat Transfer in a Free Jet against a Plane Surface

Jian-Jun SHU

School of Mechanical & Aerospace Engineering, Nanyang Technological University
50 Nanyang Avenue, Singapore 639798
Telephone: (Singapore) +65 6790 4459
Fax number: (Singapore) +65 6791 1859
E-mail address: mjjshu@ntu.edu.sg

## Abstract

A new two-layer model has been proposed to study microscale heat transfer associated with a developing flow boundary layer. As an example, a cold, microscale film of liquid impinging on an isothermal hot, horizontal surface has been investigated. The boundary layer is divided into two regions: a micro layer at microscale away from the surface and a macro layer at macroscale away from the surface. An approximate solution for the velocity and temperature distributions in the flow along the horizontal surface is developed, which exploits the hydrodynamic similarity solution for microscale film flow. The approximate solution may provide a valuable basis for assessing microscale flow and heat transfer in more complex settings.

## 1  Introduction

There has been an enormous interest in developing microscale systems. The technological world is working towards the ideology that "small is beautiful". A whole new field of studies and research has sprung up to investigate the properties of this microscale size technology. Recently noticeable progress has been made in the field of microelectromechanical systems (MEMS), thin films, superlattices and nanomaterials, it opens up the different fields of studies in the traditional engineering disciplines like heat transfer.

Heat transfer at microscale is becoming increasingly important in the development of micro-technological systems. As systems approach microscopic scale, increasing deviations from the well-established continuum laws are reported [1]. The continuum approaches such as Navier-Stokes equations are useful to describe macroscopic behavior of fluids. Fluid particles are assumed to have no mass and only translate. However, it fails to describe fluids with microstructures. The individual particle of such fluids can be of different shape and may shrink or expand. In addition, they can also rotate independently from the rotation of the fluid as a whole [2]. As the length scale of structural constituents become comparable to the intrinsic characteristic length scale of the fluid mean-free-path, the validity of the standard continuum approach with no-slip boundary conditions diminishes and sub-continuum effects are expected to become increasingly important.

In this paper the microscale heat transfer characteristics of a developing flow boundary layer is studied. As an example, the problem to be examined concerns the microscale film cooling, which occurs when a cold vertically draining sheet strikes a hot horizontal plate.

Although a sheet of fluid draining under gravity will accelerate and thin, at impact it is reasonable to model the associated volume flow as a jet of uniform velocity $U_0$ and semi-thickness $H_0$, as is illustrated in Figure 1. The notation $Q = U_0 H_0$ is introduced for the flow rate and a film Reynolds number may be defined as $R_e = \dfrac{\rho Q}{\mu}$, where $\mu$ is the dynamic viscosity of the fluid.

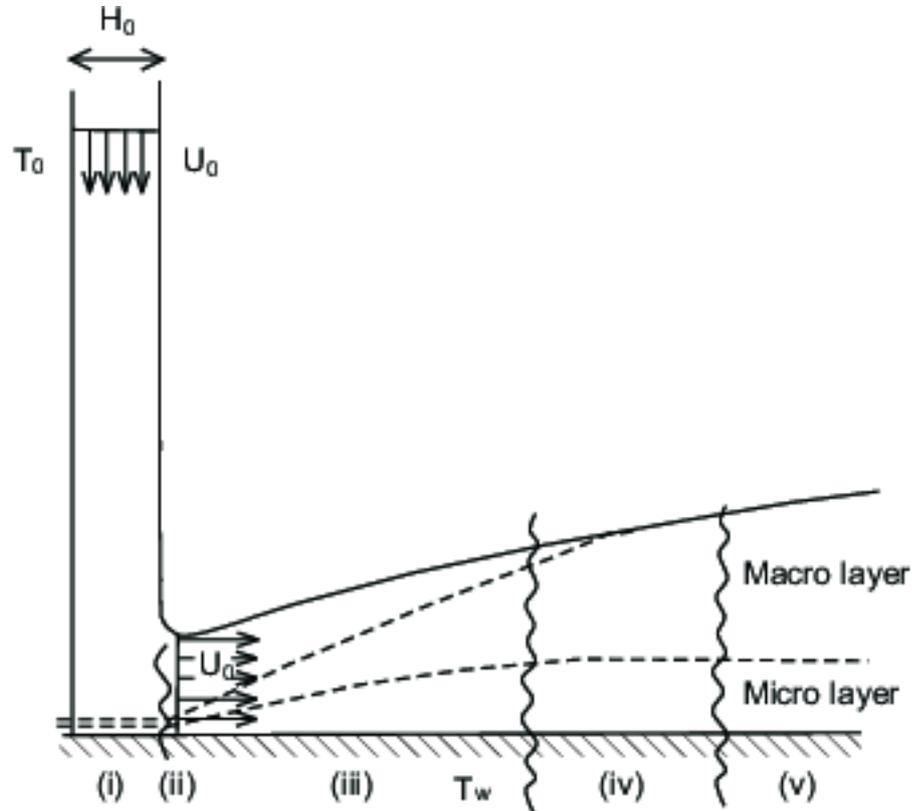

Figure 1: Flow characteristics of a two-dimensional vertical jet striking a horizontal flat plate: (i) embedded stagnation boundary layer, (ii) outer inviscid deflection region, (iii) quasi Blasius viscous diffusion, (iv) transition around viscous penetration, (v) similarity film flow.

The boundary layer has two regions: a micro layer at microscale away from the surface and a macro layer at macroscale away from the surface. The underlying hydrodynamics of the fluid flow may be summarized as [3]:
  (i) a deeply imbedded stagnation boundary layer of thickness;
  (ii) an outer inviscid deflection region, in which fluid rapidly accelerates from the value zero on the axis of symmetry to the free stream value;
  (iii) a Blasius region in which a boundary layer develops against the plane, effectively within a uniform stream;
  (iv) a transition region in which viscous effects penetrate the free surface and reduce its velocity;
  (v) a region well away from the axis of symmetry where similarity solutions for the developing film thickness, the free surface velocity and the velocity distribution can be found.

The temperature condition within which heat transfer estimates will be obtained assumes a constant temperature $T_w$ at the plane and zero heat flux at the free surface. If the rate of viscous diffusion exceeds that of temperature diffusion, the point at which viscous effects penetrate the free surface will occur before the point at which the free surface first experiences the presence of the hot plane. This physical appraisal of the developing flow field provides the framework for the initial approximate method of solution. Schematically the flow may be represented as in Figure 2 and divided into the following regions.

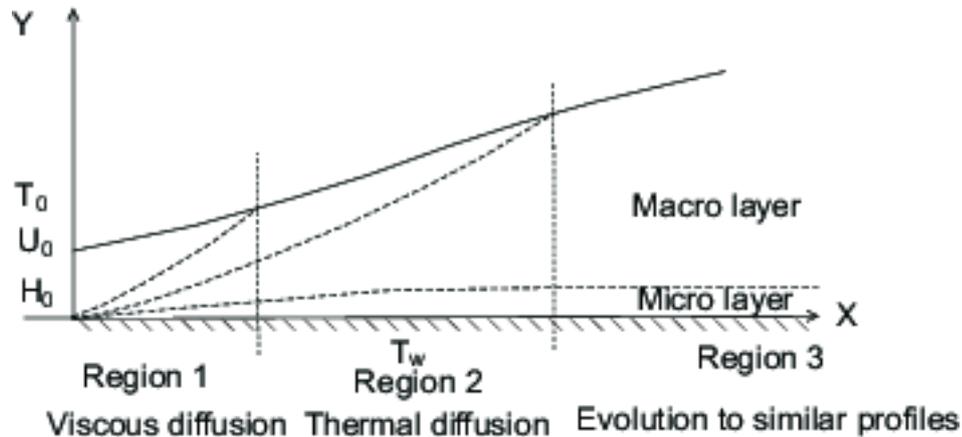

Figure 2: Basis of approximate solution.

## Region 1

In this region the impinging jet essentially experiences an inviscid symmetric division and deflection. A viscous boundary layer develops against the horizontal plate within the deflected jet and eventually penetrates the free surface marking the end of Region 1. A thermal boundary layer develops simultaneously, but for Prandtl numbers greater than unity this will still be evolving at the end of Region 1.

## Region 2

A judicious choice of approximating profiles in Region 1 is designed to approximate immediate transition to the film similarity solution at the onset of Region 2. Consequently Region 2 is examined under the assumption that the full hydrodynamic similarity solution is applicable. The adjustment of the temperature field as thermal effects develop and penetrate the free surface within this hydrodynamic setting is monitored. The end of Region 2 is notionally reached when the presence of the hot wall is first detected at the free surface.

## Region 3

In the film cooling setting, when there is zero heat flux at the free surface, the film will eventually reach a uniform temperature distribution, coinciding with the temperature of the wall. Again within the established hydrodynamics, Region 3 covers the evolution towards this asymptotic state once wall temperature effects penetrate the free surface.

## 2       Governing equations

The microscale flow under investigation has been modeled as a steady, two-dimensional flow of incompressible micropolar fluid in the micro layer and Newtonian fluid in macro layer. The two layers are distinguished by a dimensionless parameter Knudsen number $K_n$, which is defined as the ration of the fluid mean-free-path and the macroscopic length scale of the physical system. In the absence of body forces external pressure gradients and viscous dissipation the equations expressing conservation of mass, momentum and energy are consequently

**Micro layer** $(0 < Y < K_n H(X),\ T = T_w)$

$$(\mu + \mu_r)\frac{\partial^2 U}{\partial Y^2} + 2\mu_r \frac{\partial \omega}{\partial Y} = 0 \tag{1}$$

$$\gamma \frac{\partial^2 \omega}{\partial Y^2} - 2\mu_r \left(\frac{\partial U}{\partial Y} + 2\omega\right) = 0 \tag{2}$$

**Macro layer** $(K_n H(X) < Y < H(X),\ \omega = 0)$

$$\frac{\partial U}{\partial X} + \frac{\partial V}{\partial Y} = 0 \tag{3}$$

$$\rho\left(U\frac{\partial U}{\partial X} + V\frac{\partial U}{\partial Y}\right) = \mu \frac{\partial^2 U}{\partial Y^2} \tag{4}$$

$$\rho C_p \left(U\frac{\partial T}{\partial X} + V\frac{\partial T}{\partial Y}\right) = \kappa \frac{\partial^2 T}{\partial Y^2} \tag{5}$$

where $\mu_r$ is the dynamic microrotation viscosity of the fluid. In the specified physical setting, the equations are to be solved subject to the following conditions:

$$U = 0,\ \omega = -\frac{b}{2}\frac{\partial U}{\partial Y},\ T = T_w \quad \text{on}\ Y = 0,\ X \geq 0 \tag{6}$$

$$U = \varepsilon U_s,\ V = 0,\ \omega = 0,\ T = T_w \quad \text{at}\ Y = K_n H(X),\ X \geq 0 \tag{7}$$

$$\frac{\partial U}{\partial Y} = 0,\ \omega = 0,\ \frac{\partial T}{\partial Y} = 0 \quad \text{at}\ Y = H(X),\ X \geq 0 \tag{8}$$

$$\int_0^{H(X)} U(X, Y)\,dY = \text{constant} = H_0 U_0 \quad \text{for}\ X \geq 0. \tag{9}$$

In assigning variable relation between microrotation $\omega$ and skin friction $\frac{\partial U}{\partial Y}$ at the wall surface $Y = 0$, the fraction $b$ varies from zero to two. The value of $b = 0$ represents cases where the fluid particle density is sufficiently great that the microelements close to the wall are unable to rotate. The value $b = 1$ corresponds to weak concentrations, and when $b = 2$ we have cases that are representative of turbulent boundary layer.

On the assumption that the film thickness remains thin relative to a characteristic horizontal dimension, a boundary layer treatment of the equations leads to significant simplification. The following non-dimensional variables are introduced

$$x = \frac{X}{R_e H_0},\ y = \frac{Y}{H_0},\ \bar{h}(x) = \frac{H(X)}{H_0},$$

$$\bar{U} = \frac{U}{U_0}, \quad \bar{V} = \frac{R_e V}{U_0}, \quad \bar{\omega} = \frac{H_0 \omega}{U_0}, \quad \bar{\phi} = \frac{T - T_w}{T_0 - T_w}. \tag{10}$$

In the limit $R_e \to +\infty$, with $x$ remaining $O(1)$, the following equations are obtained

**Micro layer** $\left(0 < y < K_n \bar{h}(x), \bar{\phi} = 0\right)$

$$\frac{\partial^2 \bar{U}}{\partial y^2} + \lambda \frac{\partial \bar{\omega}}{\partial y} = 0 \tag{11}$$

$$\frac{\partial^2 \bar{\omega}}{\partial y^2} - \alpha \left(\frac{\partial \bar{U}}{\partial y} + 2\bar{\omega}\right) = 0 \tag{12}$$

**Macro layer** $\left(K_n \bar{h}(x) < y < \bar{h}(x), \bar{\omega} = 0\right)$

$$\frac{\partial \bar{U}}{\partial x} + \frac{\partial \bar{V}}{\partial y} = 0 \tag{13}$$

$$\bar{U} \frac{\partial \bar{U}}{\partial x} + \bar{V} \frac{\partial \bar{U}}{\partial y} = \frac{\partial^2 \bar{U}}{\partial y^2} \tag{14}$$

$$P_r \left(\bar{U} \frac{\partial \bar{\phi}}{\partial x} + \bar{V} \frac{\partial \bar{\phi}}{\partial y}\right) = \frac{\partial^2 \bar{\phi}}{\partial y^2} \tag{15}$$

where $P_r = \frac{C_p \mu}{\kappa}$ is the Prandtl number, $\lambda = \frac{2\mu_r}{\mu + \mu_r}$ and $\alpha = \frac{2\mu_r H_0^2}{\gamma}$ are the dimensionless micropolar fluid parameters. The boundary conditions now read

$$\bar{U} = 0, \quad \bar{\omega} = -\frac{b}{2} \frac{\partial \bar{U}}{\partial y}, \quad \bar{\phi} = 0 \quad \text{on} \quad y = 0, \; x \geq 0 \tag{16}$$

$$\bar{U} = \varepsilon \bar{U}_s, \quad \bar{V} = 0, \quad \bar{\omega} = 0, \quad \bar{\phi} = 0 \quad \text{at} \quad y = K_n \bar{h}(x), \; x \geq 0 \tag{17}$$

$$\frac{\partial \bar{U}}{\partial y} = 0, \quad \bar{\omega} = 0, \quad \frac{\partial \bar{\phi}}{\partial y} = 0 \quad \text{at} \quad y = \bar{h}(x), \; x \geq 0 \tag{18}$$

$$\int_0^{\bar{h}(x)} \bar{U} \, dy = 1 \quad \text{for} \quad x \geq 0. \tag{19}$$

These have been quoted in the context of the fully developed film flow field, which is approached in Region 3. These solutions provide the basis for developing comprehensive approximate solutions for the complete flow field downstream of the symmetry point of impingement incorporating Regions 1, 2 and 3.

## 3   Exact solutions for micro layer $\left(0 < y < K_n \bar{h}(x), \bar{\phi} = 0\right)$

The main differences between the governing equations of a Newtonian fluid and a micropolar fluid are firstly, the stress tensor is not symmetric in the law of conservation of momentum and secondly, micropolar fluid has an additional equation called law of conservation of angular momentum. The governing equations for linear momentum and angular momentum are:

$$\frac{\partial^2 \overline{U}}{\partial y^2} + \lambda \frac{\partial \overline{\omega}}{\partial y} = 0 \tag{20}$$

$$\frac{\partial^2 \overline{\omega}}{\partial y^2} - \alpha\left(\frac{\partial \overline{U}}{\partial y} + 2\overline{\omega}\right) = 0. \tag{21}$$

The boundary conditions can be stated as follow:

$$\overline{U} = 0, \ \overline{\omega} = -\frac{b}{2}\frac{\partial \overline{U}}{\partial y} \quad \text{on} \quad y = 0, \ x \geq 0 \tag{22}$$

$$\overline{U} = \varepsilon \overline{U}_s, \ \overline{\omega} = 0 \quad \text{at} \quad y = K_n \overline{h}(x), \ x \geq 0. \tag{23}$$

The linearly independent particular solutions for $\overline{U}$ are $\sinh(y\sqrt{\alpha(2-\lambda)})$, $\cosh(y\sqrt{\alpha(2-\lambda)})$, $y$ and $1$, while the linearly independent particular solutions for $\overline{\omega}$ are $\sinh(y\sqrt{\alpha(2-\lambda)})$, $\cosh(y\sqrt{\alpha(2-\lambda)})$ and $1$. Here $2-\lambda$ is greater than zero. This is to say that the dynamic microrotation viscosity, $\mu_r$, is always positive. This is true for most cases of liquids. Applying boundary conditions, we can obtain the velocity and angular velocity profiles, as shown below,

$$\overline{U} = \frac{\varepsilon \overline{U}_s}{4-b\lambda}\frac{y}{K_n \overline{h}}\left(4 - b\lambda \frac{y}{K_n \overline{h}}\right) + O(K_n^2) \tag{24}$$

$$\overline{\omega} = -\frac{2b\varepsilon \overline{U}_s}{(4-b\lambda)K_n}\left(1 - \frac{y}{K_n \overline{h}}\right) + O(K_n). \tag{25}$$

With Equations (24) and (25), we have successfully derives the velocity and angular velocities in terms for a micropolar fluid. They incorporate the microrotational viscosity term into the general forms of velocities and angular velocities.

## 4 Downstream similarity solutions for macro layer $(K_n \overline{h}(x) < y < \overline{h}(x), \ \overline{\omega} = 0)$

Introducing a similarity variable $\eta = \frac{y - K_n \overline{h}(x)}{(1-K_n)\overline{h}(x)}$, a stream function form of solution $\psi(x,y) = (1-K_n)\overline{U}_s(x)\overline{h}(x)f(\eta)$ leads to the similarity solution as the solution of

$$2f''' + 3c^2 f'^2 = 0,$$
$$f'(0) = \varepsilon, \ f'(1) = 1, \ f''(1) = 0. \tag{26}$$

Here $\overline{U}_s(x)$ represents the non-dimensional unknown velocity at the free surface, $c$ can be obtained analytically as

$$c = 3^{-\frac{1}{4}}\text{F}\left(\sin\frac{5\pi}{12}; \arccos\frac{\sqrt{3}-1+\varepsilon}{\sqrt{3}+1-\varepsilon}\right) = \frac{\sqrt{\pi}\Gamma\left(\frac{1}{3}\right)}{3\Gamma\left(\frac{5}{6}\right)} + \varepsilon + O(\varepsilon^2) \approx 1.402 + \varepsilon + O(\varepsilon^2)$$

where the incomplete elliptic integral of the first kind is defined as

$$\text{F}(p;\chi) = \int_0^\chi \frac{d\theta}{\sqrt{1-p^2\sin^2\theta}}.$$

The hydrodynamics are fully determined by the results

$$\overline{U}_s(x) = \frac{9c^2}{2\pi^2(x+l)}, \quad \overline{h}(x) = \frac{\pi}{\sqrt{3}(1-K_n)}(x+l). \tag{27}$$

Here $l$ is a non-dimensional shift constant reflecting that the solutions hold at large distances from the jet incidence. In due course $l$ may be estimated by further consideration of the boundary layer growth from the point of impact of the jet. With $\overline{\phi}(x,y) = \overline{\phi}(\eta)$ is readily shown that $\overline{\phi}$ satisfies

$$\overline{\phi}'' = 0, \quad \overline{\phi}(0) = \overline{\phi}'(1) = 0. \tag{28}$$

Thus, as anticipated, the asymptotic downstream solution for the temperature distribution is just $\overline{\phi}(\eta) = 0$, i.e. the temperature $T_w$ ultimately persists throughout the film if $\frac{\partial T}{\partial y} = 0$ at the free surface.

## 5  Approximate solutions

An approximate solution scheme is now presented which examines closely the flow at impingement. The solution is built up from this vicinity, stage by stage, to provide comprehensive details of the velocity and temperature distribution along the entire plate.

### 5.1 Region 1

At impact, an inviscid deflection of the draining sheet occurs over a negligibly small length scale. Essentially the flow along the plane in this region is modeled as a horizontal film of uniform velocity $U_0$ arriving at the leading edge $X = 0$ of a semi-infinite flat plate. Only after deflection will the flow be aware of the presence of the solid boundary, and only then will viscous effects begin to influence the flow field. The development of a viscous boundary layer within a uniform velocity film indicates a close parallel in this region with the Blasius boundary layer flow. Similarly the temperature differential between the plane and the fluid will only begin to influence the temperature distribution after deflection. Thus a developing thermal boundary layer may also be anticipated from $X = 0$.

The equations governing the viscous and thermal boundary layers are exactly the same as (13)-(15), but the boundary conditions now read

$$\overline{U} = \varepsilon, \quad \overline{V} = 0, \quad \overline{\phi} = 0 \quad \text{on} \quad y = K_n \overline{h}(x), \; x \geq 0$$

$\overline{U} \to 1, \; \overline{\phi} \to 1$ as $y$ approaches the outer limits of the viscous and thermal boundary layers respectively

$$\overline{U} = 1, \; \overline{\phi} = 1 \quad \text{at} \quad x = 0, \; y > 0.$$

Their solutions for $P_r > 1$ indicate that the length scale of thermal diffusion can be significantly less than that of viscous diffusion. Viscous effects, in due course, must penetrate the free surface and the transition region of Figure 1 is essentially a region of adjustment from the Blasius profile to the similarity profile. As the profiles are not greatly dissimilar, a device that in effect compresses the transition region to a single point is introduced. An approximate velocity profile

$$\overline{U}(x,y) = f'(\eta), \quad \eta = \frac{y - K_n \overline{h}(x)}{\delta(x)} \tag{29}$$

is assumed, where $\delta(x)$ is the non-dimensional boundary layer thickness. A polynomial approximation to the velocity profile is more convenient. To maintain the aggregate and matching properties of $f'(\eta)$, and simultaneously exploit the convenience of a polynomial representation, a fourth-order polynomial approximation to $f'(\eta)$ has been obtained as

$$f'(\eta) = \varepsilon + (c_0 + \varepsilon)\eta + (4 - 3c_0 + 7\varepsilon)\eta^3 + (2c_0 - 3 + 5\varepsilon)\eta^4,$$

where $c_0$ is the constant $\dfrac{\sqrt{\pi}\,\Gamma\!\left(\dfrac{1}{3}\right)}{3\Gamma\!\left(\dfrac{5}{6}\right)} \approx 1.402$. The profile is then used in a Kármán-Pohlhausen method of solution. Over Region 1 unretarded fluid is present when $x < x_0$, say where $x_0$ marks the point of penetration of viscous effects at the free surface, so that $\bar{U}_s(x) = 1$ and $\delta(x) < \bar{h}(x)$ over $0 < x < x_0$. For $x > x_0$ into Region 2 $\delta(x) \equiv \bar{h}(x)$ and $\bar{U}_s(x) < 1$ in a manner which, using the conservation of flow constraint, can be matched directly onto the asymptotic similarity solutions. The momentum integral equation reads

$$\frac{d}{dx}\int_{K_n\bar{h}(x)}^{\delta(x)+K_n\bar{h}(x)} \bar{U}(1-\bar{U})\,dy = \left(\frac{\partial \bar{U}}{\partial y}\right)_{y=K_n\bar{h}(x)} + O(\varepsilon^2) \tag{30}$$

and using (29) leads to the solution

$$\delta^2 = (19.775 + 6.898\varepsilon)x + O(\varepsilon^2) \tag{31}$$

where $\delta(x) = 0$ has been assumed at $x = 0$, which is valid in the limit of the underlying assumption. Invoking the conservation of volume flow at $x_0$, the end point of Region 1 leads to

$$\int_{K_n\bar{h}(x)}^{\delta(x)+K_n\bar{h}(x)} \bar{U}\,dy + (\bar{h} - \delta) = 1 + O(\varepsilon^2) \tag{32}$$

whence

$$\bar{h}(x) = 1 + \frac{3(4 - c_0 - 5\varepsilon)\delta}{20} + O(\varepsilon^2). \tag{33}$$

Since $\delta(x_0) = \bar{h}$

$$x_0 = 0.136 - 0.381\varepsilon + O(\varepsilon^2) \tag{34}$$

and matching the free surface velocity at $x = x_0$ leads to

$$l = 0.76 + 1.659\varepsilon + O(\varepsilon^2). \tag{35}$$

The polynomial $f'(\eta)$ is consequently used in subsequent developments of velocity and temperature distributions. It remains to establish the temperature characteristics in Region 1. The energy integral equation of (15) becomes

$$\frac{d}{dx}\int_{K_n\bar{h}(x)}^{\delta_T(x)+K_n\bar{h}(x)} \bar{U}(1-\bar{\phi})\,dy = \frac{1}{P_r}\left(\frac{\partial \bar{\phi}}{\partial y}\right)_{y=K_n\bar{h}(x)} + O(\varepsilon^2), \tag{36}$$

where $\delta_T(x)$ denotes the outer limits of the region of thermal diffusion. For $P_r > 1$, $\delta_T(x) < \delta(x)$ over $0 < x < x_0$. The notation $\eta_T = \dfrac{y - K_n\bar{h}(x)}{\delta_T(x)}$ is introduced and the ratio $\frac{\delta_T}{\delta}$ is denoted by $\Delta$ so that $\eta = \Delta \eta_T$. The solution for $\delta_T(x)$ is again developed by assuming profiles for $\overline{U}$ and $\overline{\phi}$ as

$$\overline{U}(\eta) = f'(\eta), \quad \overline{\phi}(\eta_T) = f'(\eta_T) \qquad (37)$$

which ensures identical velocity and temperature distributions for $P_r = 1$ when also $\Delta = 1$. Assuming a constant ratio $\Delta$ leads to

$$P_r \Delta^2 = \frac{0.142}{D(\Delta)} - \frac{(0.055 - 0.046\Delta - 0.021\Delta^3 + 0.011\Delta^4)\varepsilon}{D^2(\Delta)} + O(\varepsilon^2), \qquad (38)$$

where $D(\Delta) = \Delta(0.149 - 0.005\Delta^2 - 0.003\Delta^3)$. The values of $\Delta$ can obtained numerically for various Prandtl numbers. Notice that as a result of the choice of approximating profile the velocity distribution at the end of Region 1 exactly matches that of Region 2.

## 5.2 Region 2

In Region 2, the hydrodynamics are governed by the similarity solution where thermal diffusion continues to progress across the film. Accordingly the velocity at the free surface is no longer uniform, but is prescribed in non-dimensional terms by (27). The film thickness $\overline{h}(x)$ and the viscous boundary layer thickness $\delta(x)$ now coincide as

$$\delta(x) = \overline{h}(x) = \frac{\pi}{\sqrt{3}(1-K_n)}(x+l). \qquad (39)$$

The energy integral equation (36) remains appropriate. The presence of the free surface limits further viscous penetration and $\delta_T(x) \to \delta(x) = \overline{h}(x)$. In prescribing profiles $\eta_T = \frac{y - K_n \overline{h}(x)}{\delta_T(x)}$ may again be utilized, but now $\Delta(x) = \frac{\delta_T(x)}{\delta(x)}$ is no longer constant, and must in fact tend to 1 at the end of Region 2. The following profiles are introduced into the energy equation:

$$\overline{U}(x,\eta) = \overline{U}_s(x) f'(\eta), \quad \overline{\phi}(x,\eta_T) = f'(\eta_T). \qquad (40)$$

The equation for $\delta_T(x)$ is accordingly

$$\delta_T(x) \frac{d}{dx}\left\{\overline{U}_s(x) \delta_T(x) \int_0^1 f'(\eta)[1 - f'(\eta_T)] d\eta_T \right\} = \frac{c_0 + \varepsilon}{P_r} + O(\varepsilon^2). \qquad (41)$$

This resultant first-order equation in $\Delta^2$ may now be integrated with initial data $\Delta(x_0; P_r)$ as far as $\Delta(x_1(P_r); P_r) = 1$ to give

$$\Delta^2 (0.299 - 0.019\Delta^2 - 0.014\Delta^3) + \Delta(0.39 - 0.535\Delta - 0.597\Delta^3 + 0.393\Delta^4)\varepsilon$$
$$= \frac{0.476 - 0.951 K_n - 0.339\varepsilon}{P_r} \ln \frac{x+l}{x_1+l} + 0.266 - 0.349\varepsilon + O(\varepsilon^2). \qquad (42)$$

$x_1$ marks the end of Region 2, as predicted, using the polynomial profile. Beyond $x_1$ viscous and thermal effects are present throughout the film.

## 5.3 Region 3

The boundary condition of zero heat flux at the edge of the developing thermal layer in Region 1 and Region 2 is based on the assumption of a continuous temperature distribution developing smoothly into the impinging jet temperature. Once the temperature effects of the hot wall penetrate the free surface beyond $x_1(P_r)$ the zero heat flux boundary condition remains appropriate. Here, however, it reflects the insulating role of the surrounding air. As a consequence, the temperature of the film will now rise as a result of continuing heat input at the plate. In fact the temperature of the film will now progress to $T_w$, so long as the insulating boundary condition is maintained.

To accommodate the adjustment of the film temperature to $T_w$, the following profiles are adopted

$$\overline{U}(x,\eta) = \overline{U}_s(x) f'(\eta), \quad \overline{\phi}(x,\eta) = \beta(x) f'(\eta) \tag{43}$$

where now $\eta = \dfrac{y - K_n \overline{h}(x)}{\overline{h}(x)}$. The energy integral equation now reads

$$\frac{d}{dx} \int_{K_n \overline{h}(x)}^{\overline{h}(x) + K_n \overline{h}(x)} \overline{U}(\beta - \overline{\phi}) dy - \int_{K_n \overline{h}(x)}^{\overline{h}(x) + K_n \overline{h}(x)} \overline{U} \frac{d\beta}{dx} dy = \frac{1}{P_r} \left( \frac{\partial \overline{\phi}}{\partial y} \right)_{y = K_n \overline{h}(x)} + O(\varepsilon^2). \tag{44}$$

The result is an equation for $\beta(x)$ within the framework of prescribed film thickness, namely

$$(0.142 + 0.052\varepsilon) \frac{d}{dx} (\overline{U}_s \overline{h} \beta) - (0.61 + 0.75\varepsilon) \overline{U}_s \overline{h} \frac{d\beta}{dx} = \frac{1.402 + \varepsilon}{P_r} \frac{\beta}{\overline{h}} + O(\varepsilon^2) \tag{45}$$

and hence,

$$\beta(x) = \left( \frac{x_1 + l}{x + l} \right)^{\frac{1.015 - 2.03 K_n - 2.237\varepsilon}{P_r}} + O(\varepsilon^2) \tag{46}$$

which satisfies the requirements $\beta(x_1(P_r)) = 1$ and has $\beta(x) \to 0$ at rates dependent on $P_r$.

## 6    Approximate solution results

The approximate solution scheme outlined provides comprehensive details of the flow and heat transfer characteristics for the model flow. Estimates of film thickness, velocity and temperature distributions, skin friction and heat transfer coefficients over the entire region downstream of the point of impingement can be obtained.

The elements of interest in engineering practice are the shear stress at the solid boundary, *i.e.* the skin friction and the rate of heat transfer at the boundary. The skin friction is defined as

$$\tau = \mu \left( \frac{\partial U}{\partial Y} \right)_{Y = K_n H(X)} \tag{47}$$

leading to the non-dimensional skin friction coefficient

$$\overline{\tau} = \frac{H_0 \tau}{\mu U_0} = \frac{\tau R_e}{\rho U_0^2} = \left( \frac{\partial \overline{U}}{\partial y} \right)_{y = K_n \overline{h}(x)}. \tag{48}$$

The approximate solutions give

$$\begin{aligned}
\overline{\tau} &= (0.315 + 0.17\varepsilon) x^{-\frac{1}{2}} + O(\varepsilon^2) &&\text{in Region 1} \\
&= (0.693 - 0.693 K_n + 1.482\varepsilon)(x + l)^{-2} + O(\varepsilon^2) &&\text{in Regions 2 and 3}.
\end{aligned} \tag{49}$$

The integrable square root singularity is consistent with the Blasius boundary layer equivalent. The most significant film cooling design factor is the heat transfer across the film. The heat transfer at the solid boundary is given by

$$q = -\kappa \left(\frac{\partial T}{\partial Y}\right)_{Y=K_n H(X)} = \frac{\kappa \Delta T}{H_0}\left(\frac{\partial \bar{\phi}}{\partial y}\right)_{y=K_n \bar{h}(x)}, \qquad (50)$$

where $\Delta T = T_w - T_0$. The non-dimensional version is the Nusselt number defined as

$$N_u = \frac{qH_0}{\kappa \Delta T} = \left(\frac{\partial \bar{\phi}}{\partial y}\right)_{y=K_n \bar{h}(x)}. \qquad (51)$$

The results are

$$N_u = \frac{1}{\Delta(P_r)}\frac{0.315 + 0.17\varepsilon}{\sqrt{x}} + O(\varepsilon^2) \qquad \text{in Region 1}$$

$$= \frac{1}{\Delta(x; P_r)}\frac{0.773 - 0.773 K_n + 0.551\varepsilon}{x+l} + O(\varepsilon^2) \qquad \text{in Region 2} \qquad (52)$$

$$= \frac{0.773 - 0.773 K_n + 0.551\varepsilon}{x+l}\left(\frac{x_1 + l}{x+l}\right)^{\frac{1.015 - 2.03 K_n - 2.237\varepsilon}{P_r}} + O(\varepsilon^2) \qquad \text{in Region 3}.$$

The values of $\Delta(P_r)$ have been obtained from equations (38) and $\Delta(x; P_r)$ is the solution of equation (42).

## 7   Concluding remarks

An approximate and the elements of engineering practice, namely the skin friction and heat transfer coefficients for the microscale flow of a cold two-dimensional jet against a hot, horizontal plate has been presented. Although at this stage a comparison between theory and experiment is unavailable, the work provides the basis for re-assessing microscale condensation.

## References

1. Ho, C. M. and Tai, Y. C. 1998 Micro-electro-mechanical systems (MEMS) and fluid flows. *Annual Review of Fluid Mechanics*, **30**, 579-612.
2. Eringen, A. C. 2001 *Microcontinuum Field Theories II: Fluent Media*, Springer Verlag.
3. Shu, J.-J. and Wilks, G. 1996 Heat transfer in the flow of a cold, two-dimensional vertical liquid jet against a hot, horizontal plate. *International Journal of Heat and Mass Transfer* **39**(16), 3367-3379.